\documentclass[a4paper]{article}
\title{A Concept for Kilotonne Scale Liquid Argon Time Projection Chambers}

\usepackage[british]{babel} 
\usepackage[utf8]{inputenc} 
\usepackage{lmodern}
\usepackage[T1]{fontenc} 
\usepackage{textcomp}
\usepackage[svgnames]{xcolor}
\usepackage{placeins}

\usepackage[a4paper,top=3cm,bottom=2cm,left=3cm,right=3cm,marginparwidth=1.75cm]{geometry}

\usepackage{hyperref}
\usepackage{mathtools} 
\usepackage{amsfonts}
\usepackage{amssymb}
\usepackage{physics}
\usepackage[version=4]{mhchem}
\usepackage{tabu}
\usepackage{booktabs} 

\usepackage{afterpage}

\usepackage{siunitx}
\DeclareSIUnit\radlen{\text{\ensuremath{X_{\mathrm{0}}}}}
\DeclareSIUnit\clight{\text{\ensuremath{c}}} 

\usepackage[labelfont=bf]{caption} 

\usepackage{graphicx}
\graphicspath{{./Figures/}}
\usepackage{microtype}   
\usepackage{soul}

\usepackage[autostyle]{csquotes} 
\usepackage{xpatch} 

\newcommand*{\m}{\mathrm}

\begin{document}

\begin{center}
	
	{\Large \bf A New Concept for Kilotonne Scale Liquid Argon Time Projection Chambers} 
	\vspace*{1.0cm}
	\setcounter{footnote}{0}  
	\def\A{\kern+.6ex\lower.42ex\hbox{$\scriptstyle \iota$}\kern-1.20ex a}
	\def\E{\kern+.5ex\lower.42ex\hbox{$\scriptstyle \iota$}\kern-1.10ex e}
	\small
	\newcommand{\Aname}[2]{#1}
	\def\titlefoot#1{\vspace{-0.3cm}\begin{center}{\bf #1}\end{center}}
	
	\Aname{M.~Auger}{Bern},
	\Aname{R.~Berner}{Bern},
	\Aname{Y.~Chen}{Bern},
	\Aname{A.~Ereditato}{Bern},
	\Aname{D.~Goeldi\footnote{Now at Department of Physics, Carleton University, Ottawa, Ontario, K1S 5B6, Canada}}{Bern},
	\Aname{P.~P.~Koller}{Bern},
	\Aname{I.~Kreslo}{Bern},
	\Aname{D.~Lorca}{Bern},
	\Aname{T.~Mettler}{Bern},
	\Aname{F.~Piastra}{Bern},
	\Aname{J.~R.~Sinclair\footnote{Corresponding author: james.sinclair@lhep.unibe.ch}}{Bern},
	\Aname{M.~Weber}{Bern}, and
	\Aname{C.~Wilkinson\footnote{Corresponding author: callum.wilkinson@lhep.unibe.ch}}{Bern}
	\titlefoot{Albert Einstein Center for Fundamental Physics, Laboratory for High Energy Physics, University of Bern, 3012 Bern, Switzerland\label{Bern}}

    \Aname{M.~Convery}{SLAC},
    \Aname{L.~Domine}{SLAC},
    \Aname{F.~Drielsma}{SLAC},
    \Aname{R.~Itay}{SLAC},
    \Aname{D.~H.~Koh}{SLAC},
    \Aname{H.~A.~Tanaka}{SLAC},
    \Aname{K.~Terao}{SLAC},
    \Aname{P.~Tsang}{SLAC}, and
    \Aname{T.~Usher}{SLAC}
    \titlefoot{SLAC National Accelerator Laboratory, Stanford University, Menlo Park, California, USA\label{SLAC}}
          
	\Aname{D.~A.~Dwyer}{LBNL},
	\Aname{S.~Kohn}{LBNL},
	\Aname{P.~Madigan}, and
	\Aname{C.~M.~Marshall}{LBNL}
	\titlefoot{University of California and Lawrence Berkeley National Laboratory, Berkeley, CA 94720, USA\label{LBNL}}
	
	\Aname{A.~Bross}{FNAL}
	\titlefoot{Fermi National Accelerator Laboratory, Batavia, IL 60510, USA\label{FNAL}}
	
	\Aname{J.~Asaadi}{Arlington}
	\titlefoot{University of Texas at Arlington, Arlington, TX 76019, USA\label{Arlington}}
\end{center}
\vspace*{1cm}

\begin{abstract}
We develop a novel approach for a Time Projection Chamber (TPC) concept suitable for deployment in kilotonne scale detectors, with a charge-readout system free from reconstruction ambiguities, and a robust TPC design that reduces high-voltage risks while increasing the coverage of the light collection system. This novel concept could be deployed as a Far Detector module in the Deep Underground Neutrino Experiment (DUNE) neutrino-oscillation experiment.
For the charge-readout system, we use the charge-collection pixels and associated application-specific integrated circuits currently being developed for the liquid argon (LAr) component of the DUNE Near Detector design, ArgonCube.
In addition, we divide the TPC into a number or shorter drift volumes, reducing the total voltage used to drift the ionisation electrons, and minimising the stored energy per TPC. 
Segmenting the TPC also contains scintillation light, allowing for precise trigger localisation and a more expansive light-readout system.
Furthermore, the design opens the possibility of replacing or upgrading components.
These augmentations could substantially improve reliability and sensitivity, particularly for low energy signals, in comparison to a traditional monolithic LArTPCs with projective charge-readout.

\end{abstract}

\section{Introduction} \label{sec:Intro}

Liquid Argon Time Projection Chambers (LArTPCs) evolved from gaseous TPCs~\cite{TPC,LArIonize}, and the design of modern LArTPCs for neutrino detection~\cite{CAVANNA20181} has followed a consistent approach since Ref~\cite{LArTPC}.

The focus of the ArgonCube collaboration has been on developing a new approach to LArTPCs for maximal reliability, as well as detector sensitivity. 
ArgonCube emerged from the earlier ArgonTube~\cite{argontube} collaboration which studied the technical issues related to operating LArTPCs with long drift distances (\SI{5}{\metre}) and high electric fields (\SI{1}{\kilo\volt\per\centi\metre}).
Maximal reliability and sensitivity is achieved through detector modularisation, pixelated charge readout, and innovative light-detection modules. 
Modularisation simplifies electric field stability requirements, reduces the stored energy per TPC region, and lowers the requirements for LAr purity.
Pixelated charge readout has the potential to provide true 3D imaging of particle interactions by removing the reconstruction ambiguities present for current wire readout techniques where multiple 2D images must be combined to infer a 3D image. 
New light detection methods could increase photon yields and provide improved localisation of light signals.

To date, each technology of this design has been demonstrated individually in small-scale test-stands located at the University of Bern, Lawrence Berkeley National Lab (LBNL), and Fermilab.
The next ArgonCube prototype, to be deployed in the NuMI beam at Fermilab~\cite{numi-beam}, is a mid-scale test-bed combining all aforementioned technologies in a modular detector with a \SI[product-units=repeat]{1.4x1.4x1.2}{\metre} active volume. This will also serve as a prototype for the full ArgonCube deployment in the full DUNE Near Detector (ND). Successful demonstration of these technologies within the existing ArgonCube R\&D program has the potential to demonstrate their viability for a multi-kilotonne scale LArTPC.

In this document we present a general conceptual design for a multi-kilotonne scale LArTPC. The Deep Underground Neutrino Experiment (DUNE)~\cite{DUNE} is a planned long-baseline neutrino-oscillation experiment which is currently in development. Four Far Detector (FD) modules, each with \SI{10}{\kilo\tonne} (\SI{17}{\kilo\tonne}) of liquid argon (LAr) in the fiducial (total) volume, are envisaged~\cite{Far_Detectors}. Three of the modules are planned to be single-~\cite{single} or dual-phase~\cite{dual} Time Projection Chambers (TPCs), but the design of the TPC for the fourth module has yet to be decided. In Section~\ref{sec:DUNE}, before concluding, we provide a high-level overview of why the improvements discussed here would be suitable and advantageous for the DUNE fourth far detector module design, as a concrete application of the general design.

\FloatBarrier
\section{Pixelated Charge Readout}
Since their evolution from gaseous TPCs~\cite{TPC,LArIonize,LArTPC}, the charge readout for LArTPCs has been achieved with two or more projective readout planes, the majority of which use wire readout. Projective readouts have been successfully demonstrated in a number of experiments~\cite{icarus,argonute,uboner}, but they introduce intrinsic ambiguities in event reconstruction~\cite{ambiguous}. 
These ambiguities are the result of trying to reconstruct complex 3D objects from a limited number of 2D projections. Ambiguities are particularly problematic if tracks are aligned parallel to the readout plane, or if multiple events arrive at the same time.
To address these issues, the ArgonCube collaboration proposes a novel pixelated charge-readout system which provides 3D information about the event, where the pixelated charge-readout system is formed of charge-collection pads on a conventional Printed Circuit Board (PCB).

The approach described here evolved from initial studies at Bern~\cite{pixels} which employed an analogue signal multiplexing scheme to make use of existing wire readout electronics~\cite{larasic}. 
A scaled-up version of the Bern readout was deployed in the LArIAT TPC~\cite{lariat} for test-beam studies. 
While these tests were successful in demonstrating pixels to be a viable alternative to projective wire readout (as previously deployed in LArIAT), the analogue multiplexing scheme caused ambiguities in the event reconstruction.
To provide a true 3D readout, dedicated pixel-readout electronics were required. In answer to this need, a custom cryogenic-compatible Application-Specific Integrated Circuit (ASIC), called LArPix~\cite{larpix}, was developed at LBNL, which collects, amplifies and digitises charge from every pixel individually, and is mounted directly on the reverse of the pixel PCB. Digitising and multiplexing the signals in the cold drastically reduces the number of signal feedthroughs required, and allows each pixel to be read out completely independently, thus producing unambiguous 3D information. 

\begin{figure}[tbp]
	\centering
	\includegraphics[width=.97\textwidth]{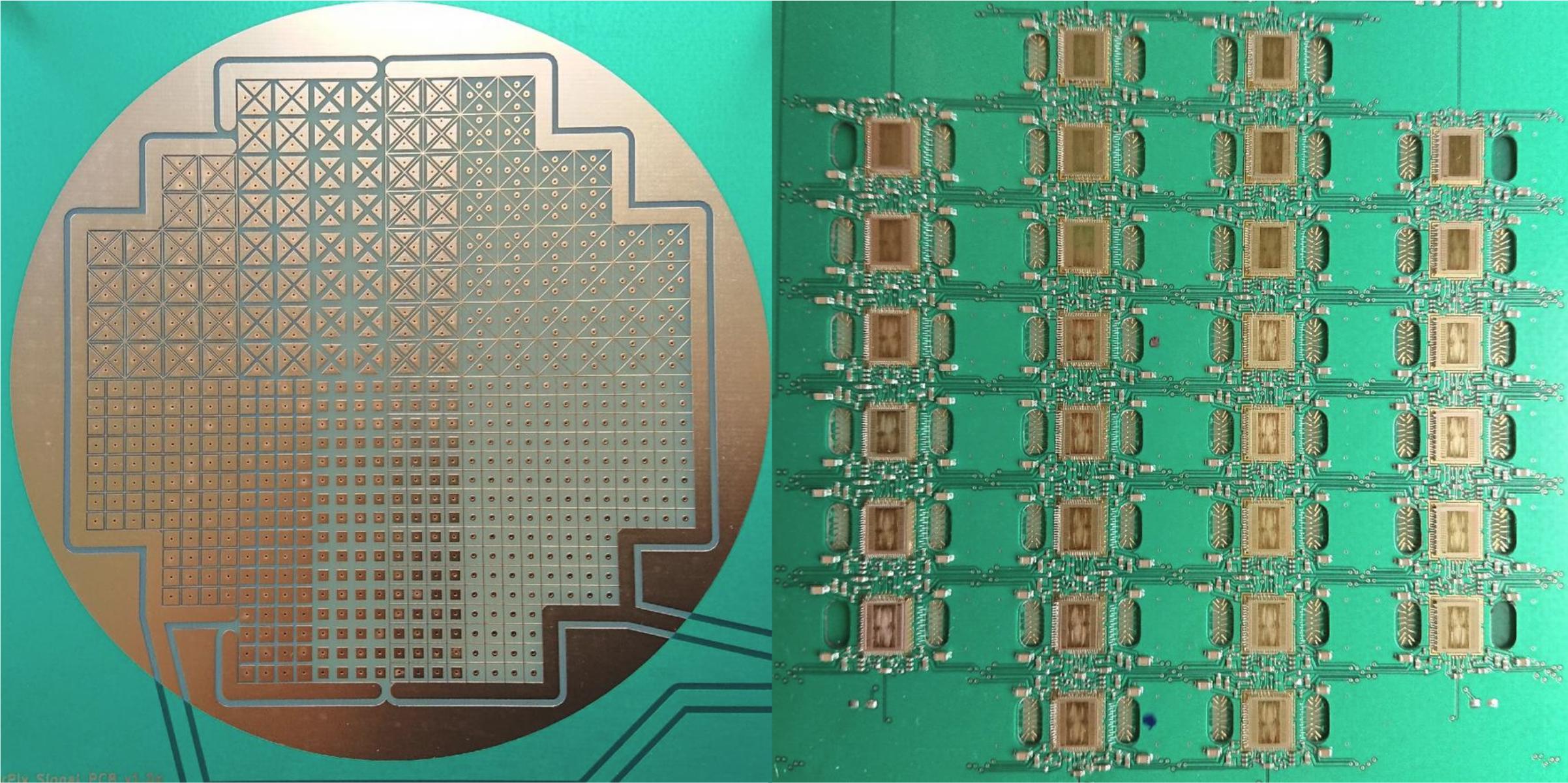}
	\caption{A prototype pixelated charge-readout PCB developed as part of the ArgonCube R\&D programme. The PCB has 832 pixels with various pad geometries, to identify optimal dimensions (left). LBNL's LArPix ASICs are mounted directly on the rear of the PCB, providing cold signal digitisation (right).}
	\label{fig:pixel_plane}
\end{figure}

\begin{figure}[htb]
  \centering
  \includegraphics[width=.29\columnwidth]{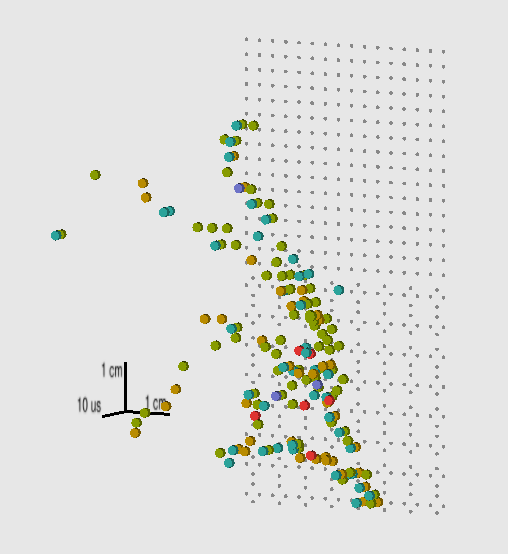}
  \includegraphics[width=.382\columnwidth]{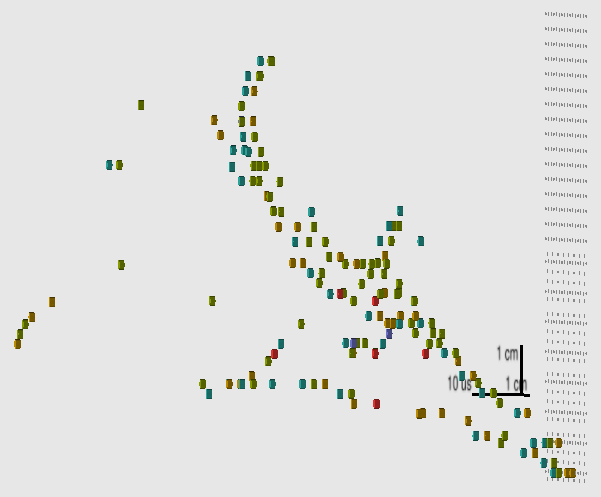}
  \includegraphics[width=.294\columnwidth]{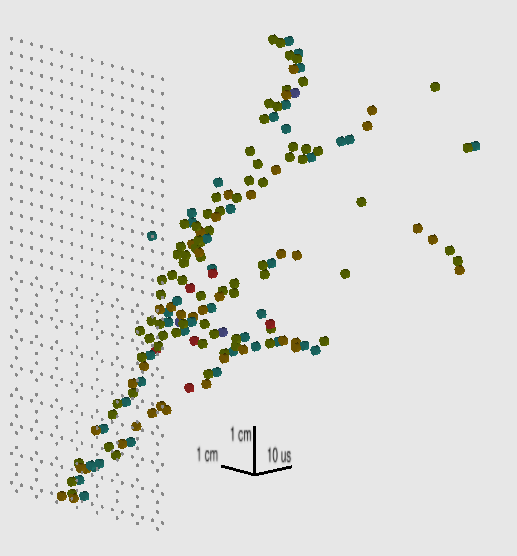}	
  \caption{A cosmic ray induced shower recorded with the LBNL LArPix equipped charge-readout system shown in Figure~\ref{fig:pixel_plane} in the Bern pixel demonstration LArTPC. The same 3D event is shown from three different angles.}
  \label{fig:pixel_results}
\end{figure}

The prototype pixel readout shown in Figure~\ref{fig:pixel_plane} was used to successfully measure the 3D ionisation distributions of cosmic rays passing through a small LArTPC demonstrator at Bern. 
The \SI{60}{\centi\metre} drift (\SI{10}{\centi\metre} diameter) demonstrator LArTPC is a charge-readout test-stand with comparable drift length and electric field, \SI{1}{\kilo\volt\per\centi\metre}, as the design proposed for the DUNE Near Detector.
Figure~\ref{fig:pixel_results} shows a cosmic induced shower recorded in the pixel demonstrator, from three different angles which can be easily interpreted without the need for any complex reconstruction, thus demonstrating the full 3D imaging capability of the pixel design.

The conventional PCB construction makes the pixelated readout mechanically stable and scalable in modular ``tiles''. The compact and modular nature of the pixel tiles means that, unlike wires which require large structures to be held in tension\footnote{For example, the \SI{12}{\centi\metre} thick Anode Plane Assemblies~\cite{single} of the DUNE single-phase FD module design.}, the tiles are potentially removable from the detector volume with minimal mechanical requirements.
This would enable upgrades of, or repairs to, individual pixel tiles without incurring significant detector downtime.

An important consequence of the full 3D LArPix charge-readout system is the lack of a preferred direction, giving a far more uniform reconstruction efficiency as a function of particle direction, unlike for wire readout TPCs. For wire readout systems, particles travelling parallel to the wire orientation are reconstructed with a much lower efficiency~\cite{ambiguous, Antonello:2012hu}. A more uniform charge reconstruction efficiency would be advantageous for many potential applications of LArTPCs, including applications where there is no preferred direction, such as proton decay or supernova-neutrino detection~\cite{SupernovaDetection}, or where the angular dependence of events is important to understand, for example solar-neutrino detection~\cite{SuperKSolar,SuperKSolar2,Solar}, or reconstructing the particles coming out of a beam neutrino interaction.

The challenge to build a low-power pixel-based charge readout for use in LArTPCs has independently inspired other research groups. We mention, in particular, the Q-Pix~\cite{Nygren:2018rbl} approach. This proposed scheme aim to capture waveforms of arbitrary complexity from a sequence of varying time intervals, each of which corresponds to a fixed charge integral.  

\subsection{Power Consumption}

Providing a unique front-end channel for each pixel was a crucial step in the development of a true 3D LArTPC readout.
Readout densities of $\sim\,$10$^5$ channels per square meter places stringent requirements on the design of the electronics.
Results with the first LArPix prototype, tested at \SI{77}{\kelvin}, have shown a total power consumption of \SI{62}{\micro\watt} (\SI{37}{\micro\watt} digital component) which corresponds to \SI{6.7}{\watt\per\metre\squared}.
An average power consumption of $\mathcal{O}\left(10\right)\,\mathrm{W\,m}^{-2}$ is tolerable, provided sufficient cooling from LAr circulation.   
However, it is important to avoid any excessive heating of the LAr to prevent localised boiling at the readout plane. 
For this reason, a conservative limit of $\leq$\SI{100}{\micro\watt} per channel has been set~\cite{larpix}.

\subsection{Data Acquisition Requirements}

One might assume that the increased channel number of a pixelated charge readout, in comparison to a projective wire readout covering the same area, would also increase the Data AcQuisition (DAQ) requirements.
However, increased channel density does not necessarily increase bandwidth demands, given the inherently sparse nature of LArTPC data.     
Pixel pitches in the range of \SIrange{3}{5}{\milli\metre} correspond to pixel densities of between 111k and 40k pixels\,m$^{-2}$, respectively.
The internal channel number scales with the pixel density.
Due to the daisy-chaining of LArPix, the external channel number (I/O lines at feedthrough) corresponds to $\mathcal{O}\left(10\right)\,\mathrm{m}^{-2}$ of readout plane. 

For a given particle interaction in a LArTPC, the vast majority of channels have no signal. Even in the intense DUNE ND environment we would expect a pixel occupancy at less than the percent level~\cite{DUNECDRVol2}. LArPix implements pixel-level self-triggering, only digitising and reading out the small fraction of channels with actual signal. A digital data acquisition rate of $\mathcal{O}\left(0.5\right)\,\mathrm{MB}\,\mathrm{s}^{-1}\,\mathrm{m}^{-2}$ is expected when exposed to the flux of surface cosmic rays with a metre-scale drift length. The low external channel density and low data rate minimises the requirements of the DAQ, for which commercially available Field-Programmable Gate Array (FPGA) boards are more than sufficient.
 
\subsection{Manufacturing Technique and Costs}

Pixel tiles of \SI[product-units=repeat]{32x32}{cm} and a populated thickness less than \SI{10}{\milli\metre} can easily be produced via commercial PCB manufacturing techniques, and tiles of this size are currently being tested as part of the ongoing ArgonCube program.
Other manufacturing techniques and costs would have to be considered for the cabling and mechanical support structures for the pixel tiles.
Design costs for a LArPix ASIC suitable for a large scale LArTPC will be minimal, assuming the ongoing design effort for the ArgonCube design for the DUNE ND reaches completion.

With total readout planes at a scale $\mathcal{O}\left(1000\right)\,\mathrm{m}^{2}$, the pixel-tile production cost is estimated to be no more than \$5k $\mathrm{m}^{-2}$ for fully commercial ASIC and PCB production and assembly.
The costs for support structures, cabling, and feedthroughs are still unknown, but are likely to be less than the pixel tiles.
Quality assurance and control costs have also yet to be determined; this work would likely be split between institutions and industrial partners.  

\section{Reducing High Voltage Requirements}

\begin{figure}[htbp]
  \centering
  \includegraphics[width=.8\columnwidth]{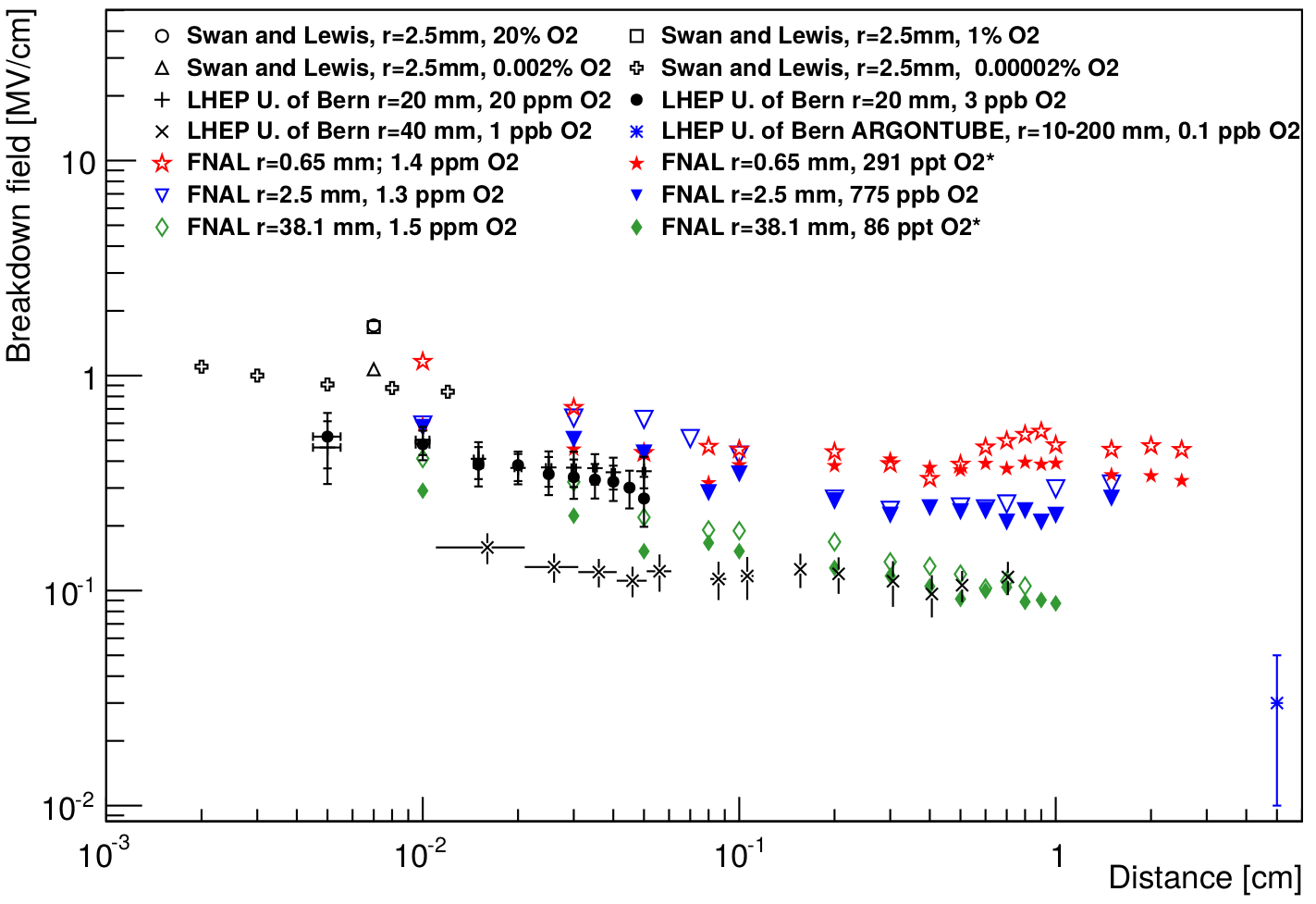}\\
  \includegraphics[width=.8\columnwidth]{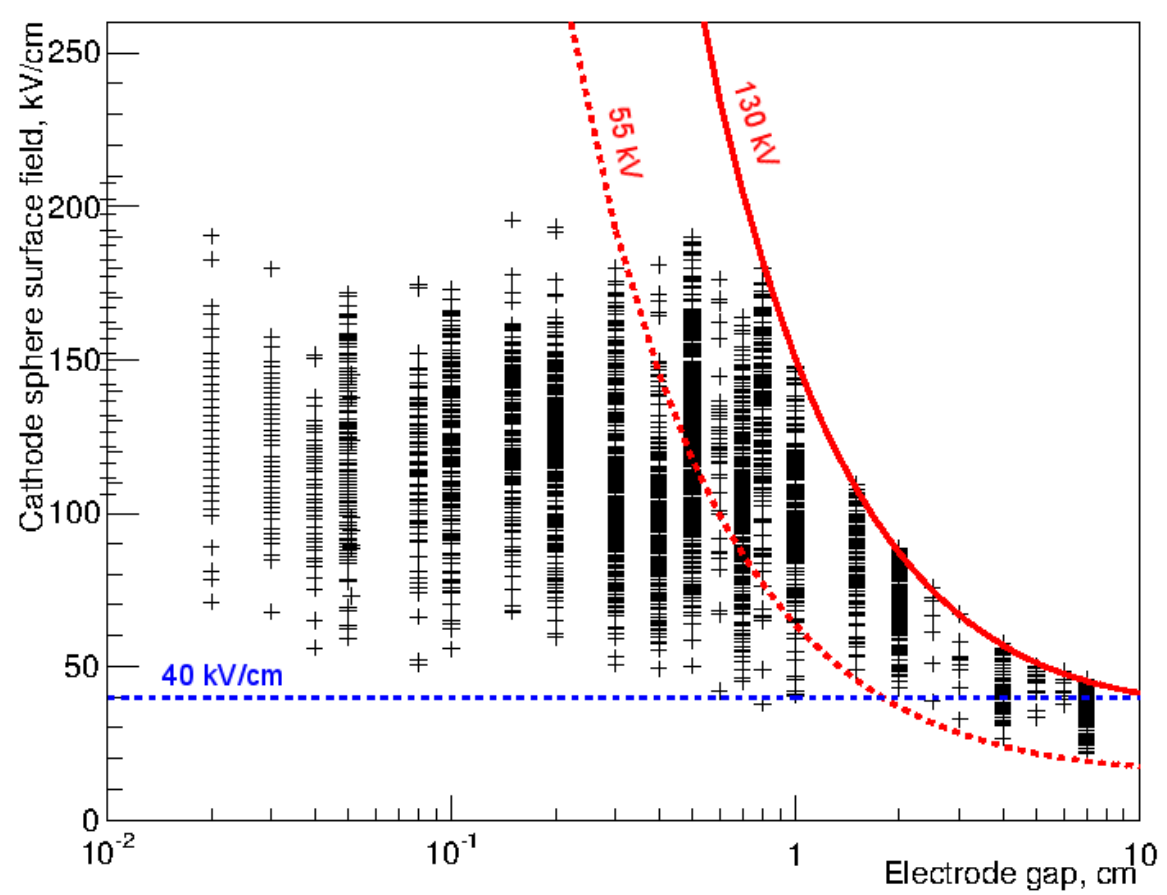}
  \caption{Top, a comparison of all dielectric breakdown measurements in liquid argon, showing breakdowns at centimetre scale~\cite{FIAL_HV}. Bottom, breakdown point in liquid argon; the electric field at a spherical cathode is plotted against the anode to cathode separation (electrode gap)~\cite{HVoriginal}.}
  \label{fig:breakdown}
\end{figure}
For LArTPCs operated at high voltages, there is a risk that small imperfections within the TPC and surrounding volume can lead to damaging breakdowns. Such issues have prevented many liquid noble gas TPCs from achieving their target voltages. Studies have shown that breakdowns can occur in electric fields as low as \SI{40}{\kilo\volt\per\centi\metre} in LAr~\cite{FIAL_HV,HVoriginal}, see Figure~\ref{fig:breakdown}. As ArgonTube~\cite{argontube_design} demonstrated, it is extremely difficult to reach and maintain high drift fields at large drift distances, and the reasons are often not well understood.

There are two possible solutions to the problem. Firstly, one can shorten the drift length, which reduces the high cathode voltage requirements and reduces LAr purity requirements~\cite{lngDet}. As in ICARUS~\cite{ICARUS_detector}, the drift length could be split into two separate TPCs which share a common central cathode, reducing the required voltage by a factor of two. Such a division can be implemented multiple times in order to reach a desired safe operating cathode voltage. Secondly, reducing detector capacitance is important in order to reduce the amount of stored energy in the TPC that could be released in the event of a breakdown. This can be reduced by segmenting the cathode planes into multiple sections with each cathode section electrically isolated from its neighbours. This prevents a HV breakdown from affecting the entire drift volume, as each cathode segment supports an independent TPC (or a pair of TPCs for a shared central cathode), and reduces the stored energy in each TPC module.

\subsection{Field Shell}
Accommodating a large number of independent TPCs inside a large cryostat capable of supporting a multi-kilotonne LAr volume requires a novel approach to electric-field shaping. Dead material must be minimised in order to maximise the active volume. We propose replacing traditional field-shaping rings with a continuous resistive plane forming a ``field-shell''.
This would provide a continuous linear potential distribution along the drift direction, paired with simple mechanics.
By eliminating the resistor chain, the component count is reduced drastically and therefore also the potential points of failure.
In the case of a breakdown, a resistive plane would limit the rate of energy dissipation.

For a material to be suitable for use as a field-shell, it must have a uniform resistance of $\mathcal{O}\left(10\right)\,\mathrm{G}\Omega\,\mathrm{sq}^{-1}$ at the desired voltage and temperature, \SI{-70}{\kilo\volt} and \SI{87}{\kelvin}.
The field-shell can be produced with solely resistive material in the form of a $\mathcal{O}\left(100\right)\,\mu\mathrm{m}$ foil, or as a laminate with a $\mathcal{O}\left(10\right)\,\mu\mathrm{m}$ resistive layer on a G10 substrate\footnote{Glass fibre reinforced resin~\cite{g10}.}.
The electro-magnetic radiation length of G10, $X_{\m{0}} = \SI{19.4}{\centi\metre}$, and  its hadronic interaction length, $\lambda_{\m{int}} = \SI{53.1}{\centi\metre}$, are both comparable to LAr, \SI{14}{\centi\metre} and \SI{83.7}{\centi\metre}, respectively~\cite{pdg_g10}.
This means that including G10 structures in LAr does not introduce significant complications to the interpretation of the detector response relative to a monolithic design.
A G10 substrate would also provide a strong dielectric, capable of \SI{200}{\kilo\volt\per\centi\metre} at \SI{1}{\centi\metre} thick~\cite{G10Breakdown},  layer around the field-shell.
This dielectric shielding eliminates the need for a clearance volume between the TPCs and the cryostat, while also shielding the TPC from breakdowns in a neighbouring TPC.
The field shell also keeps the power consumption low: for a TPC of unit volume (\SI[product-units=repeat]{1x1x1}{\metre}) at a field of \SI{0.5}{\kilo\volt\per\centi\metre} and a field shell with a resistance per square of $4~\mathrm{G}\Omega\,\mathrm{sq}^{-1}$, the power consumption would be $\sim$\SI{2.5}{\watt}.

\begin{figure}[htb]
  \centering
  \includegraphics[width=.4\textwidth]{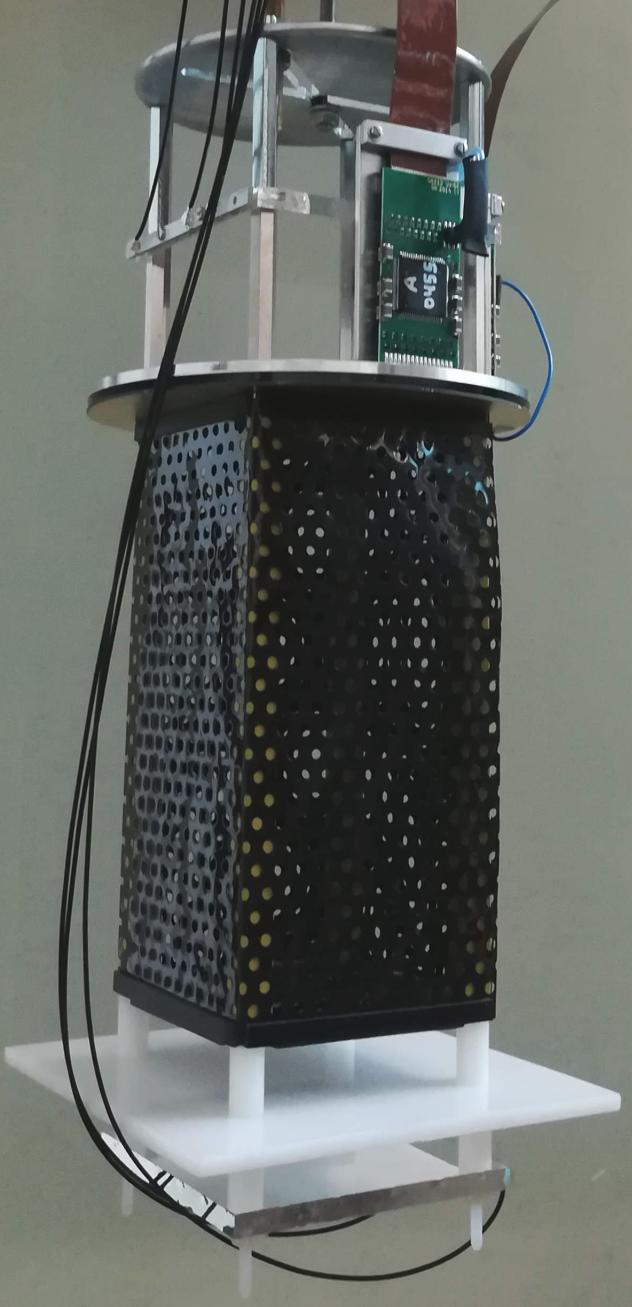}
  \caption{The Bern field-shell demonstrator LArTPC. 
    The field-shell and cathode are made from $\sim\,$\SI{50}{\micro\metre} resistive kapton foil. 
    For this test, the field-shell was perforated to allow purification of the LAr within the active volume.
    The TPC has a \SI[product-units=repeat]{7x7}{\centi\metre} footprint and a \SI{15}{\centi\metre} drift length. 
    A bias of up to \SI{-23}{\kilo\volt} was applied to the cathode to generate a \SI{1.5}{\kilo\volt\per\centi\metre} drift field~\cite{FSTPC}.}
  \label{fig:shellTPC}
\end{figure}
A field-shell demonstrator LArTPC at Bern, see Figure~\ref{fig:shellTPC}, has shown the technology to be viable~\cite{FSTPC}.
The TPC has a \SI[product-units=repeat]{7x7}{\centi\metre} footprint and a \SI{15}{\centi\metre} drift length, with the field-shell constructed from a single sheet of $\sim\,$\SI{50}{\micro\metre} carbon-loaded Kapton\footnote{DuPont\texttrademark, Kapton\textregistered \, polyimide film, E. I. du Pont de Nemours and Company, \url{www.dupont.com}.} foil.
The TPC was exposed to cosmic activity, triggering on crossing muons, at electric fields up to \SI{1.5}{\kilo\volt\per\centi\metre}. 
It operated successfully, with straight tracks observed across a range of electric fields.  

\section{Optical Segmentation}

To minimise the stored energy, we have proposed segmenting the cathode and using field-shells to isolate individual TPCs, forming TPC modules. 
A further consequence of a modular detector volume is that the scintillation light is contained within each TPC module.
The prompt scintillation light, $\tau<$ \SI{6.2}{\nano\second}~\cite{scintillation}, can be efficiently measured with improved timing resolution, provided a dielectric light readout with $\mathcal{O}\left(1\right)\,\mathrm{ns}$ timing resolution, such as ArCLight~\cite{arclight}, is deployed within the field-shells. The cost for build an ArCLight-type system is driven by the cost of wave-length shifting plastic, and is $\sim$\$10.5k $\mathrm{m}^{-2}$

Optical segmentation in a short drift allows us to use a large-area optical detection system such as is outlined in (ArCLight). Although the photon detector efficiency in ArCLight is $\sim$1\%, this means that the threshold for detecting a light signal is $\sim$\SI{50}{\kilo\electronvolt}.
This also would allow for more efficient and accurate association of light to charge signals in an environment where otherwise multiple signals could pile-up in the same drift window. Examples of pile-up include multiple beam-induced events in an intense neutrino beam such as will be sampled at the DUNE ND, or a large supernova-neutrino signal in a large detector.
Fast timing enables the association of energy deposits to the correct interaction vertex, regardless of spatial separation of deposit from the vertex.   

Additionally, optical segmentation combined with high-efficiency optical detection could provide complementary calorimetry from the scintillation light. Intrinsic scintillation light should give you nanosecond level timing. Rayleigh scattering can cause a degradation of timing resolution, depending on the intensity of the light signal. This could reduce sensitivity to fast processes such as kaon decay, particularly when the light intensity is low.

\section{Deployment as a DUNE Far Detector Module}
\label{sec:DUNE}

Although the improvements to the LArTPC design presented above are general to any multi-kilotonne LArTPC deployment, it is instructive to take a specific example. The Deep Underground Neutrino Experiment (DUNE)~\cite{DUNE} is a planned long-baseline neutrino-oscillation experiment which is currently in development, with kilotonne scale LArTPCs in use as Far Detector (FD) modules (each of which contain \SI{17}{\kilo\tonne} of LAr)~\cite{Far_Detectors}. There is a trade-off between drift-length, segmentation, readout performance and cost, which would require sophisticated study to settle on a final design.

In the case of a DUNE FD module, the benefits to the pixelated charge readout come from the more uniform readout efficiency as a function of particle direction afforded by LArPix. Pixelated charge-readout would improve the energy estimation of beam neutrinos as high-angle particles would not be missed, and would improve the background detection efficiency, both of which are important for DUNE's long-baseline neutrino oscillation programme. It would particularly improve the reconstruction of low energy energy events which have no preferred direction, such as proton decay and supernova-neutrino detection~\cite{SupernovaDetection}, which are key pillars of DUNE's physics programme, and additionally, make the reconstruction of solar-neutrinos~\cite{SolarDUNE} more uniform as a function of zenith-angle, which benefit the study of angular dependent effects~\cite{SuperKSolar,SuperKSolar2,Solar}.

The benefits to a DUNE FD module from the segmented optical readout proposed here come from the containment of prompt scintillation light, which enables a precise and localised trigger, negating the need for any dead time during, or after, high-energy events occurring elsewhere in the detector. Furthermore, the contained scintillation light would allow for a more efficient detection of the total light yield for a single event, improving particle identification and, in turn, reducing backgrounds. These characteristics would improve the search for solar- (\textless \SI{10}{\mega\electronvolt}) and supernova-neutrinos (\SIrange{10}{30}{\mega\electronvolt}), as well as the proton-decay channel $p\rightarrow K^{+} + \bar{\nu}$, where sensitivity to the 12.4 ns kaon lifetime is vital in distinguishing it from the atmospheric neutrino background.   
 
Segmenting perpendicular to the drift direction increases the number of optical readout planes and readout systems required, as well as increasing the amount of dead material. Although, the optical readout out planes can be regarded as active, in that a significant signal will be provided for any crossing charged particle.   

It has been remarked that the use of a continuous resistive field shell laminated onto a G10 substrate provides dielectric shielding between the neighbouring TPCs, and additionally negates the need for a dielectric clearance volume between the TPCs and the cryostat, potentially expanding the active and therefore the fiducial volumes for a common cryostat design\footnote{The clearance volume for the proposed single-phase DUNE FD modules is \SI{20}{\centi\metre} thick~\cite{single}.}. It has also been remarked that the EM radiation and hadronic interaction lengths are similar between G10 and LAr, meaning that the dead material introduced is not too problematic for reconstruction.

\begin{figure}[tbp]
  \centering
  \includegraphics[width=.7\textwidth]{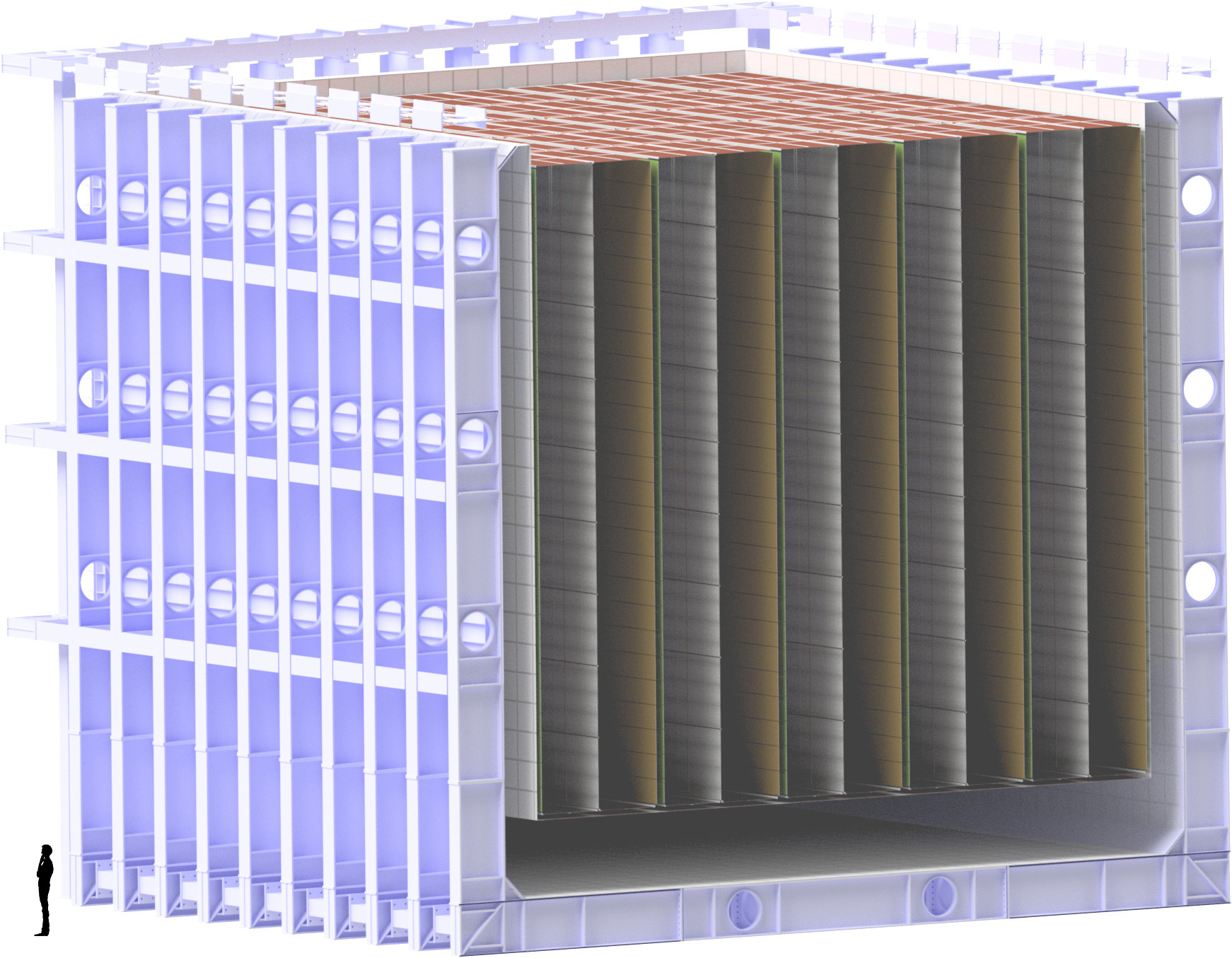}
  \caption{Potential deployment of the modular LArTPC concept discussion in this work in a DUNE far detector cryostat. The internal cryostat dimensions are \SI{62}{\metre} long, \SI{15.1}{\metre} wide and \SI{14}{\metre} high. A figure is shown in the bottom left to give a sense of scale. There are five cathodes which divide the volume vertically into 10 separate TPCs along the drift axis (shown as white panels), with 6 corresponding charge-readout pixel anode planes (shown as brown panels). Optical readout planes would enclose the face which is cut away, on both sides of the TPC. The division of the cathodes into 20 sections along the beam axis (into the figure) is visible along the top face of the detector. All 200 individual TPCs are contained within a G10 structure within the cryostat.}
  \label{fig:dune_fd}
\end{figure}

Although we stress that a detailed optimisation would be required, a potential suitable segmentation scheme for a DUNE FD module, which have internal cryostat dimensions \SI{62}{\metre} long (in the beam direction), \SI{14}{\metre} tall, and \SI{15.1}{\metre} wide~\cite{DUNECDRVol3}, is shown in Figure~\ref{fig:dune_fd}. Note, this scheme aims to reduce HV risks, but is not necessarily optimized for light collection. In this design, we retain a \SI{10}{\centi\metre} clearance volume at the sides and \SI{20}{\centi\metre} at the top bottom of cryostat to accommodate infrastructure. The volume is therefore \SI{61.8}{\metre} long, \SI{13.6}{\metre} tall, and \SI{14.9}{\metre} wide. With ten TPCs across the width, there are five shared \SI{5}{\milli\metre} thick cathodes, and six \SI{30}{\milli\metre} thick pixel anode planes. The drift length for each TPC is therefore \SI{1.47}{\metre}. The length is segmented into twenty independent sections, with \SI{5}{\milli\metre} G10 sheets forming the field shells, and \SI{10}{\milli\metre} thick optical readout planes lining each vertical wall of the field shell. Each \SI{3.09}{\metre} cathode section would be electrically isolated from its neighbours, and would support an independent pair of TPCs, preventing a HV breakdown from affecting the entire detector volume. The active volume in our design is $\sim$21\% larger than for the DUNE single-phase modules~\cite{Far_Detectors}\footnote{The DUNE single-phase active volume is given as \SI{58}{\metre} long (in the beam direction), \SI{12}{\metre} tall, and \SI{14.5}{\metre} wide~\cite{Far_Detectors}.}, although fiducialisation studies have not been carried out, and are complicated by the high degree of segmentation.

Reducing the drift length increases the number of anode planes, and therefore the total cost of the charge-readout electronics, as well as the quantity of dead material in the detector. However, for a total drift length of \SI{15.1}{\metre} divided into ten TPCs in our design, with a conservative assumption of \SI{30}{\milli\metre} thick pixel anode planes, the six anode planes would correspond to only \SI{18}{\centi\metre} of dead material\footnote{By comparison, the DUNE anode plane assemblies are \SI{12}{\centi\metre} thick, and there are three of them across the drift direction, giving \SI{36}{\centi\metre} of dead material~\cite{single}.}.

In this scheme, with a \SI{1.47}{\metre} drift length, only \SI{-73.5}{\kilo\volt} is required at each cathode to maintain an equivalent field to the current single-phase DUNE FD design\footnote{The DUNE single-phase FD module design has a central cathode, and requires a \SI{-180}{\kilo\volt} to maintain a field of \SI{500}{\volt\per\centi\metre}.}, of \SI{500}{\volt\per\centi\metre}. It has been shown in Ref.~\cite{AT} that a LArTPC can be operated at this voltage without a prohibitive loss in active volume. By approximating a TPC as a parallel-plate capacitor, it is possible to calculate the combined stored energy for a pair of TPCs sharing a common cathode segment in this scheme. For a TPC with a cathode \SI{13.6}{\metre} tall and \SI{3.09}{\metre} wide, and a \SI{1.47}{\metre} drift length, the capacitance is \SI{425}{\pico\farad}. At \SI{-73.5}{\kilo\volt}, the combined (two TPCs per cathode) \SI{850}{\pico\farad} corresponds to a stored energy\footnote{The stored energy per cathode segment of the DUNE single-phase FD module design is $\sim$\SI{100}{\joule}~\cite{Far_Detectors}.} of only \SI{2.2}{\joule}.

For the design presented here, the readout costs are not prohibitive, despite the increase in readout area due to the segmented design. The pixelated charge readout cost is predicted to be $\sim$\$5k~$\mathrm{m}^{-2}$. In order to instrument a DUNE FD module with the charge-readout proposed here, the total cost would be $\sim$\$42M, which scales with the number of TPCs along the width (ten in this case). The optical readout cost is predicted to be $\sim$\$10k~$\mathrm{m}^{-2}$. In order to instrument a DUNE FD module with the optical-readout proposed here, the total cost would be $\sim$\$80M, which scales with the number of segmentations along the beam axis (twenty in this case).

\section{Summary}
\label{sec:summary}

In this work, we have outlined a new concept for LArTPCs, suitable for deployment at multi-kilotonne scales. The key enhancements are based on the ArgonCube collaboration's extensive R\&D experience, which has been aimed at the design of the LArTPC for the DUNE near detector complex. The main features are pixelated charge-readout with cold digitisation electronics free of ambiguities which naturally allows for full 3D tracking; segmentation of the detector to reduce the high voltage requirements, and mitigate the potential damage due to energy discharge in a TPC segment as a result of breakdowns in the detector; an optical system which benefits from the segmentation of the detector to contain scintillation light, and improve the precision with which a vertex can be localised and used as a trigger; and a continuous resistive shell in place of a field cage, to reduce the number of components and therefore points of failure with respect to field-shaping rings, and to reduce the amount of dead material introduced as a result of detector segmentation. Our robust modular design, with charge- and optical-readout isolated between modules offers the possibility of upgrading or replacing parts of the detector without significant downtime. The predicted cost for the charge- (optical-)readout proposed here is $\sim$\$5k~$\mathrm{m}^{-2}$ ($\sim$\$10k~$\mathrm{m}^{-2}$).

As an example, the concept has been applied to the case of one of the four expected DUNE far detector modules, a planned experiment using multi-kilotonne LArTPCs. Although the exact design for this, or any other purpose depends on an sophisticated optimisation of the drift-length, longitudinal segmentation, readout performance and cost, we show that our design could enhance many aspects of the DUNE physics programme. As well as the charge-readout improving the high angle performance and background rejection of the DUNE neutrino long-baseline oscillation programme, the improved light collection system has the potential to increase sensitivity to low energy events in various parts of DUNE's physics programme (proton decay, solar- and supernova-neutrino events). Additionally, the reduced HV requirements reduce the field strength required to produce the same detector performance, and drastically reduce the stored energy which could be released in a breakdown.

\section*{Acknowledgements}
The Bern group is supported by the Swiss National Science Foundation grant 200021-169045 and the Canton of Bern, Switzerland.
The LBNL group is supported by the Department of Energy under contract DE-AC02-05CH11231.
The UTA group is supported by the Department of Energy under contracts DE-SC0011686 and DE-SC0017721.
A. Bross is supported by Fermi Research Alliance, LLC under Contract No. DE-AC02-07CH11359 with the U.S. Department of Energy, Office of Science, Office of High Energy Physics.
SLAC is managed by Stanford University under DOE/SU Contract DE-AC02-76SF00515.
\bibliography{main}
\bibliographystyle{JHEP.bst}

\end{document}